\documentstyle[aps,preprint,epsf]{revtex}
\title{Statistical Properties of Inelastic Lorentz Gas}

\author{{\sc Kim}
Hyeon-Deuk \footnote{E-mail:kim@yuragi.jinkan.kyoto-u.ac.jp}
  and Hisao {\sc Hayakawa}\footnote{E-mail: hisao@yuragi.jinkan.kyoto-u.ac.jp}}

\address{Graduate School of Human and Environmental Studies, Kyoto University, Kyoto 606-8501}

\begin{document}
\sloppy
\maketitle

\begin{abstract}
{The inelastic Lorentz gas in cooling states is studied. 
It is found that the inelastic Lorentz gas is
localized and that the mean square displacement of the inelastic Lorentz
gas obeys a power of a logarithmic function of time.  
It is also found that the scaled position distribution of the
inelastic Lorentz gas has an exponential tail, while the distribution is
 close to the Gaussian near the peak. 
Using a random walk model, we
derive an analytical expression of the mean square displacement as a
 function of time and the restitution
coefficient, which well agrees with the data of our simulation. 
The exponential tail of the scaled position distribution function is
also obtained by the method of steepest descent. }
\end{abstract}

\newpage

\section{Introduction}
The statistical properties of dissipative systems have created
considerable attention from the view point of non-equilibrium
statistical mechanics. 
The analysis of gas kinetics has played an important role in the
construction of statistical mechanics. 
Similarly, granular gases which are collections of inelastic particles\cite{fluid,review}
are expected to be important to understand non-equilibrium statistical
mechanics of dissipative systems. 
To understand the statistical properties of dissipative
systems, several attempts have been carried out on a collection of
inelastic particles. 

The simplest model of granular gases is an inelastic hard core model. 
To analyze the time evolution of the velocity distribution function
(VDF), we usually adopt the inelastic Boltzmann equation which is a
generalization of the Boltzmann equation to the case of inelastic
gases. 
There have been several investigations on VDF of the
inelastic gas by the inelastic Boltzmann
equation.\cite{Brito,Noije,Espiov,Brey2,Brey1,Jose,Brill,Chapmann} 
The approximate analytical
solution of the inelastic Boltzmann equation has been 
obtained by the method of expansion by the Sonine
polynomial which is known as a fast convergent one and becomes standard
technique to describe both non-uniform elastic
gases and uniform inelastic gas
systems\cite{Brito,Noije,Espiov,Brey2,Brey1,Jose,Brill,Chapmann}. 
They also suggested that the scaled 
VDF of the inelastic Boltzmann
equation has an exponential high energy
tail\cite{Brito,Noije,Espiov,Brey2}, which has been confirmed by the direct simulation Monte Carlo method (DSMC)\cite{Brey2}
or in molecular dynamics simulations of smooth hard inelastic disks in
the homogeneous cooling state (HCS)\cite{Brito}. 
On the other hand, some investigations on the spatial structure of the
inelastic Boltzmann gas system have been
performed\cite{Espiov,Brey3,Chen}. 
Esipov and
P$\mathrm{\ddot{o}}$schel\cite{Espiov} have suggested that the diffusion
length of particles obeys a power of a logarithmic function of time. 
Brey et al.\cite{Brey3} have derived the
self-diffusion coefficient by means of a Chapmann-Enskog
expansion in the density gradient of the particles in the inelastic
Boltzmann gas system. 
Chen et al.\cite{Chen} have shown in the molecular dynamics simulation
of sticky inelastic particles that the probability density function of 
inelastic particles obeys an exponential distribution at late stages. 

There is, however, a strong restriction in the applicability of the
inelastic Boltzmann equation and the method of the Sonine expansion. 
Inelastic collisions generate string-like clusters in a cooling state
which sometimes reaches an inelastic collapse state. 
Thus, HCS cannot be kept in general
\cite{Pierre,Gold,Mc,Chen}. 
Thus, both of the inelastic Boltzmann equation and the method of
the Sonine expansion which are only valid in dilute
HCS\cite{Brito,Noije,Brey2,Brey1,Jose,Brill,Chapmann} become no
longer valid as time goes on\cite{Norman}. 

On the other hand, the inelastic scatterings of
classical electrons by ions can be modeled by the inelastic
Lorentz gas\cite{Edwards,Martin} in which few fraction of particles can move freely
among many immobile scatters known as Drude's formula.\cite{Jackson}   
Since the fixed scatters keep their initial
configuration, the
system can keep the state of dilute mobile particles if the initial
configuration is dilute. 
Furthermore, there are some examples of the inelastic
Lorentz gas in the natural
phenomena such as the spontaneous interparticle
percolation.\cite{Edwards,Martin} 
The Lorentz gas is also familiar in dairy life such as
the pedestrian motion among a cloud of persons and obstacles,
the pachinko game and the pinball game\cite{Luw} etc. 
There is an additional advantage that the analysis of the inelastic
Lorentz gas model is easier than that of the inelastic Boltzmann equation
because the inelastic Lorentz gas model treats the
one-body problem: The
situation of Lorentz gas is dilute mobile particles among almost immobile
scatters, while Boltzmann gas is a collection of mobile particles. 

We believe that the systematic studies of the inelastic
Lorentz gas system is also needed.  
The steady VDF of inelastic Lorentz gas and its percolation
velocity under the uniform
external field have been analyzed.\cite{Edwards,Martin}
However, there have been no efforts to investigate the spatial structure
of the inelastic Lorentz gas system without external field. 
It is important to discuss the spatial structure of the
inelastic Lorentz gas without any external field as the first step of
the study of the inelastic Lorentz gas system. 
Thus, we intend to investigate the spatial structure of the inelastic
Lorentz gas in cooling state in this paper. 

The organization of this paper is as follows. 
In $\S$ 2, we introduce the model of our simulation. 
In $\S$ 3, we validate our simulation code by
evaluating the mean square displacement, the velocity distribution and
the position distribution in elastic collisions. 
We also explain the
main results of our simulation and analysis for the mean square
displacement and the position distribution of the inelastic Lorentz
gas in $\S$ 3. 
We conclude our results in $\S$ 4. 
 
\section{The model of our simulation}
In our simulation we adopt the event driven method for hard core
collision of particles.\cite{see,Allen,event} 
The system consists
of 40000 fixed smooth hard disks, 
that is 40000 scatters, and one tracer disk.
The collision between the tracer and the scatters is inelastic hard core 
collision with a constant coefficient of restitution $e$. 
Through our simulation, we use dimensionless quantities in
which the scales of length and velocity are respectively
measured by the diameter of the disks $d$ and the standard deviation 
of the initial velocity distribution $U$. 
Namely, both of all the scatters and the tracer are circular disks with
their diameter $d$. 
The number density of the scatters in this system is
$0.04d^{-2}$, where the system size is
$L_{x}=L_{y}=1000d$. 
Because the area fraction is approximately $0.0314$, the system of our
simulation is dilute enough.   
We adopt the periodic boundary condition in 
this system. 
The system size is enough large to remove the boundary
effect of this system, because the tracer is localized and 
have rarely gone across the boundaries. 
We average each result over 10000
samples which consist of 100 different initial
configurations of 
the scatters and 100 different initial velocities
of the tracer. 
Each initial configuration of the scatters is at random. 
Each tracer starts from the center of
the system and its distribution of the initial velocities is made from
normal random numbers, where its average is
zero and its standard deviation is $U$. 
Each sample has been calculated until 100 inelastic
collisions. 

\begin{figure}[htbp]
\epsfxsize=10cm
\centerline{\epsfbox{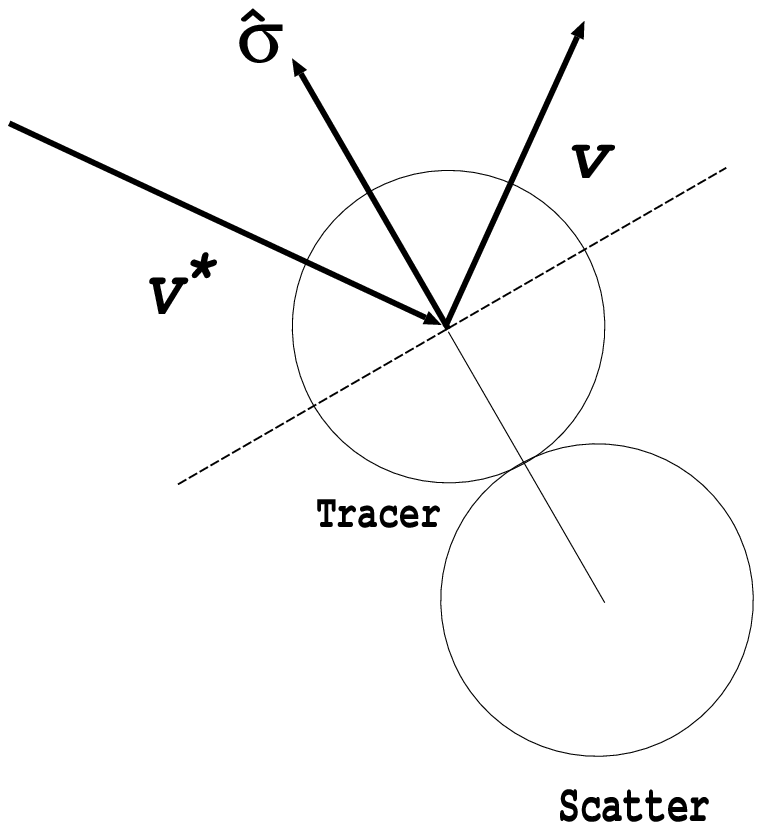}} 
\caption{Unit vector ${\bf \hat{\sigma}}$ pointing from the center of a scatter to the
center of a tracer. 
The precollision velocity of the tracer and the postcollision velocity
 of it are expressed as ${\bf v}^{*}$ and ${\bf v}$ , respectively. }
\label{collision}
\end{figure}

In each inelastic collision, the velocity
of a tracer tangential to the collision plane are preserved, while the
velocity normal to it changes with the restitution coefficient
 $e$ which is assumed to be a constant for all the
collisions. 
The precollision velocity of the tracer ${\bf v}^{*}$ 
yields the postcollision velocity of it ${\bf v}$ with
${\bf v} \cdot {\bf \hat{\sigma}}=-e ({\bf v}^{*}
  \cdot {\bf \hat{\sigma}})$, where ${\bf \hat{\sigma}}$ is a
unit vector pointing from the center of the scatter to the
center of the tracer ( Fig.\ref{collision} ) . 
Thus, the expression is given by
\begin{equation}
{\bf v}={\bf v}^{*}-(1+e)({\bf v}^{*} \cdot {\bf
  \hat{\sigma}}){\bf \hat{\sigma}}.\label{vchange}
\end{equation} 
This relation between the precollision velocity and the postcollision
velocity in the inelastic Lorentz gas system is quite different from that 
in the inelastic Boltzmann gas system. 

In the simulation of the inelastic Lorentz gas, the velocity 
of the tracer quickly decreases. 
For the sake of long time simulation, we adopt the rescaling 
of the speed of the tracer as follows. 
After each collision, the magnitude of the postcollision
velocity $|{\bf v}|$ has been put to that of the precollision
velocity $|{\bf v}^{*}|$ , i.e. the magnitude of the initial
velocity of the tracer has been preserved for all the
collisions in each sample. 
Note that we use non-scaled values of ${\bf v}$ the magnitude of which
decreases by collisions for our analysis. 

\section{Results: simulation and theory}
Now, we explain the main results of our simulation and analysis for
inelastic Lorentz gas. 
We study the mean square displacement and the position
distribution of the tracer. 

\subsection{Check of our simulation code}
Our simulation code has been validated by the test
simulation in which we have evaluated the mean
square displacement, the velocity distribution and the position distribution of
the tracer in the elastic collisions, 
i.e. $e=1$. 
For elastic scatters, we expect that the tracer displays
random walk motion when the initial speeds of the tracer are
identical and that the Gaussian is the stable velocity
distribution. 

Corresponding to a random walk model, we have fixed the
magnitudes of 100
different initial velocities of the tracer, i.e.
$|{\bf v}_{0}|=U$. 
In each sample, we calculate the motion of tracer until 10000 elastic collisions. 
The other conditions of the test simulation are same as
those of the simulation in inelastic collisions. 
We have checked that the mean square displacement of the
tracer $\langle{\bf r}^{2}(t)\rangle$ is almost a linear function of
time, where $\langle\cdot\cdot\cdot\rangle$ indicates the average over the 10000 samples. 
We also confirm that all the data of the position distribution $g(r,t)$
can be scaled as 
\begin{equation}
g(r,t) = \frac{1}{\sqrt{\langle{\bf
    r}^{2}(t)\rangle}} G(R),  \label{gtilde}    
\end{equation}
where $G(R)$ is the scaled position distribution
function and $R\equiv r/\sqrt{\langle{\bf r}^{2}(t)\rangle}$ is
the scaled distance from the center of the system at which
the tracer started. 
We check that the scaled position distribution $G(R)$ is
almost on the Gaussian curve as expected from the
conventional random walk.
Furthermore, the velocity
distribution of the tracer becomes almost the Gaussian. 

\subsection{Mean square displacement}
Let us discuss the statistical properties of inelastic Lorentz gas. 
At first, we evaluate the time evolution of the thermal velocity and
discuss the mean square displacement in the inelastic system. 

In the inelastic Lorentz gas system, many inelastic
collisions happen at random. 
The average number of collision times per unit time is evaluated as 
$2dv_{0}( t)\rho_{0}$, where $d$ is the diameter of the tracer and the
scatters and $\rho_{0}$ is the number density of the scatters.  
The thermal velocity $v_{0}( t)$ is defined as 
\begin{equation}  
v_{0}( t) \equiv  \sqrt{\frac{1}{n} \int {\bf v}^{2}f({\bf
  v},t)\mathrm{d}{\bf v}},\label{v0defi} 
\end{equation}  
with the velocity distribution of the tracer $f({\bf
  v},t)$ and the average density of the tracer $n \equiv
\int f({\bf v},t)d{\bf v}$. 
Therefore the temperature of the system is denoted as
\begin{equation}
T( t)=\frac{1}{2}mv_{0}^{2}( t),\label{thermo}
\end{equation}
where $m$ is the mass of the tracer.\cite{Brito,Noije,Brill,Brey2}

In each collision, the energy loss of the tracer is
considered to be proportional to the temperature of the system $T( t)$. 
Thus, we can estimate the energy loss of the tracer per
unit time as
\begin{equation}
\frac{\mathrm{d}\mathit{T}( t)}{\mathrm{d}\mathit{t}}=-2dv_{0}( t)\rho_{0}h( e) T( t),\label{tdiff}
\end{equation}
where $h( e)$ is a dimensionless function of the restitution coefficient
$e$. 
$h( e)$ represents a cooling rate of the inelastic Lorentz gas system. 
Substitution of eq. (\ref{thermo}) into eq. (\ref{tdiff}) leads
to 
\begin{equation}
\frac{\mathrm{d}\mathit{v}_{\mathrm{0}}\mathit{( t)}}{\mathrm{d}\mathit{t}}=-d\rho_{0}h( e) v_{0}^{2}( t).\label{vdiff}
\end{equation}
Solving eq. (\ref{vdiff}) , we obtain the thermal velocity
$v_{0}(t)$ as
\begin{equation}
v_{0}( t)=\frac{1}{\frac{1}{v_{0}(0)}+d\rho_{0}h( e)
  t} ,\label{veq}
\end{equation}
where $v_{0}(0)$ is the initial thermal velocity. 
In the inelastic Boltzmann gas system, Haff \cite{Haff} has derived the same expression as that of
eq.(\ref{veq}), where the cooling rate is assumed to be proportional to
$1-e^{2}$.   

\begin{figure}[htbp]
\epsfxsize=12cm
\centerline{\epsfbox{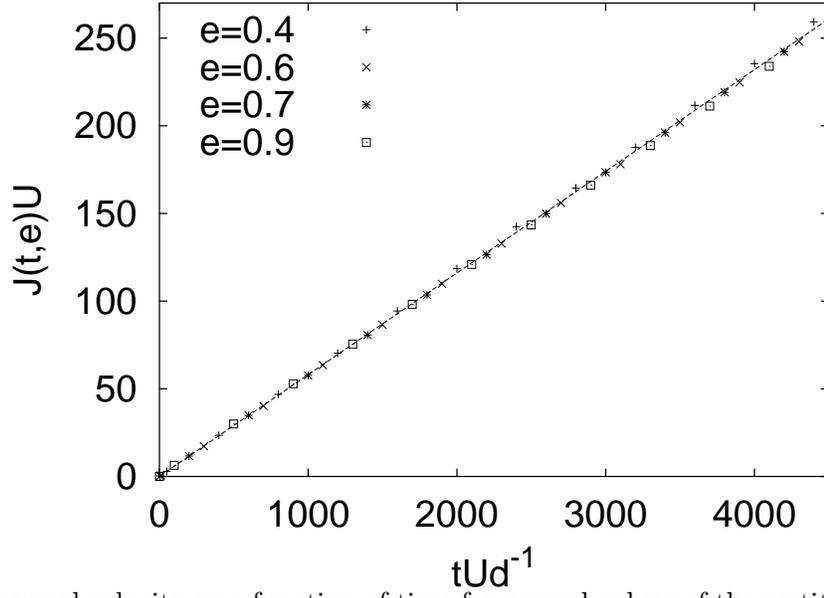}} 
\caption{Thermal velocity as a
 function of time for several values of the restitution
 coefficient $e$. Note that $J(t,e)\equiv (v_{0}^{-1}( t)-v_{0}^{-1}(0))/(1-e)$. The straight line $0.058\times tUd^{-1}$ shown as the broken line is the
 best fitting function of the results of our simulation. }
\label{v0re}
\end{figure}

Here, we can deduce the expression of $h( e)$ from our simulation
as follows. 
Figure \ref{v0re} shows $(v_{0}^{-1}( t)-v_{0}^{-1}(0))/(1-e)U$
versus $tUd^{-1}$ plots for several
values of the restitution coefficient, namely, for
$e=0.4, 0.6, 0.7$ and $0.9$. 
All the data are on the same straight line which is also shown in
Fig.\ref{v0re}. 
Corresponding to the results of our simulation, we set the number density
of the scatters $\rho_{0}=0.04d^{-2}$ and the initial thermal
velocity $v_{0}(0)=\sqrt{2}U$ from eq.(\ref{v0defi}). 
Thus, Fig.\ref{v0re} indicates that $h
( e)$ is a linear function of the restitution
coefficient $e$ as 
\begin{equation}
h( e)\simeq 1.45 \times (1-e),  \label{hsimu} 
\end{equation}
whereas Martin and Piasecki\cite{Martin} have assumed energy losses at
collisions in inelastic Lorentz gas system are proportional to
$1-e^{2}$. 
This Haff's law, $h(e)\sim 1-e^2$, can be derived from the Gaussian VDF. 
We will demonstrate that the cooling rate of the inelastic Lorentz gas
system is not proportional to 
$1-e^{2}$ and close to eq.(\ref{hsimu}) in terms of non-Gaussian nature
of VDF as follows. 

The time evolution of VDF for inelastic 
Lorentz gas without any external field is in general described by a rate
equation. 
For elastic gases, the rate equation is known as the Boltzmann equation. 
Similarly we can derive a rate equation of VDF for inelastic Lorentz gas 
systems (see Appendix A), i.e. the inelastic Lorentz gas equation : 
\begin{equation}
\frac{\partial f({\bf v},t)}{\partial t}= d \rho_{0}\int_{{\bf v} \cdot {\bf \hat{\sigma}}>0} ({\bf v} \cdot {\bf \hat{\sigma}}) [\frac{f({\bf v}^{*},t)}{e^{2}}-f({\bf v},t)]  \mathrm{d}{\bf \hat{\sigma}}. \label{inelaslorentz2}
\end{equation}
We assume that eq.(\ref{inelaslorentz2}) admits an scaling
solution depending on time only through the thermal velocity $v_{0}( t)$
defined in eq.(\ref{v0defi}) as 
\begin{equation}
f({\bf v},t)=\frac{n}{v_{0}^{2}( t)} F({\bf c}), \label{tdepend}
\end{equation} 
with the scaled velocity of a tracer ${\bf c}\equiv {\bf v}/v_{0}
( t)$ and the average density of the tracer $n\equiv \int f({\bf v},t)
d{\bf v}$.  
$F({\bf c})$ is the scaled VDF of the
tracer which is independent of time. 
Here we define the dimensionless collision integral as
\begin{equation}
I(F)\equiv \int_{{\bf c} \cdot {\bf \hat{\sigma}}>0} ({\bf c} \cdot {\bf \hat{\sigma}}) [\frac{F({\bf c}^{*})}{e^{2}}-F({\bf c})]  \mathrm{d}{\bf \hat{\sigma}}, \label{if}
\end{equation}
and its moments as 
\begin{equation}
M_{p} \equiv - \int c^{p} I(F)  \mathrm{d}{\bf c}. \label{mup}
\end{equation}
Multiplying eq.(\ref{inelaslorentz2}) by $m{\bf v}^{2}/2$ and integrating
by ${\bf v}$, we obtain the rate of change of the temperature from
eqs. (\ref{v0defi}), (\ref{thermo}), (\ref{if}), and (\ref{mup}) as 
\begin{equation}
\frac{\mathrm{d}\mathit{T}( t)}{\mathrm{d}\mathit{t}}=- d v_{0}( t) \rho_{0}  M_{2} T( t).\label{tdiff2}
\end{equation}
Comparing eq.(\ref{tdiff2}) with eq. (\ref{tdiff}), we obtain 
\begin{equation}
h(e)=\frac{M_{2}}{2}.\label{compare2}
\end{equation}
Both sides of eq.(\ref{compare2}) represent the cooling rate of the
inelastic Lorentz gas system. 
When we adopt the lowest order Sonine expansion in eq.(\ref{sonine1}),  
the expression of
$M_{2}$ as a function of the restitution coefficient $e$ is given as
\begin{equation}
M_{2}=\sqrt{\pi}(1-e^{2})(\frac{65}{64}+\frac{15}{32}e^{2}-\frac{1}{64}\sqrt{-383+828e^{2}+900e^{4}}).\label{compare3}
\end{equation}
The details of the derivation of eq.(\ref{compare3}) are written in
Appendix B.  
Figure \ref{hetomu2} shows the both sides of eq.(\ref{compare2}), 
i.e. $M_{2}/2$ from eq.(\ref{compare3}) and $h(e)$ from
eq.(\ref{hsimu}) which has been estimated from the data of our
simulation, as a function of the restitution coefficient $e$. 
There are no significant differences between $ M_{2}/2 $ and $h(e)$,
though $M_{2}/2$ does not exist in the high inelastic region. 
This means that Haff's law, $h(e)\sim 1-e^2$, may be replaced by
eq.(\ref{hsimu}) in the inelastic Lorentz gas system as a result of
non-Gaussian nature. 

Now, let us discuss the statistical properties of the inelastic Lorentz
gas based on a random walk model. 
To simplify the expressions, we introduce $\alpha$ and $\beta$ as, 
\begin{eqnarray}
\alpha \equiv \frac{1}{v_{0}(0)},\quad \beta \equiv d\rho_{0}  h(e). \label{ab}
\end{eqnarray}
Using $\alpha$ and $\beta$, the average time interval $\tau_{N}$ from $N-1$th
collision to $N$th collision between the tracer and the
scatters is given as, 
\begin{equation}
\tau_{N} \simeq \frac{l}{v_{0}( t_{N-1})}=l(\alpha +\beta 
t_{N-1}), \label{taun}
\end{equation}
where $l$ denotes
the mean free path of the tracer, and $t_{N-1}$ is the total 
time until the $N-1$th collision. 
From the relation $t_{N-1} = \tau_{1}+\tau_{2}+\cdot \cdot \cdot+\tau_{N-1}$, we obtain $\tau_{N}$ as a function of the number
of collision times $N$, 
\begin{equation}
\tau_{N} = l\alpha (1+l\beta )^{N-1}. \label{taun}
\end{equation}
Therefore the total time until the $N$th collision $t_{N}$
is given by 
\begin{equation}
t_{N} = \tau_{1}+\tau_{2}+\cdot \cdot \cdot+\tau_{N} = \frac{\alpha }{\beta }  \{(1+l\beta )^{N}-1\} \label{tn}. 
\end{equation}
Rearranging eq.(\ref{tn}), the number
of collision $N$ is obtained as a function of the
total time until the $N$th collision $t_{N}$ as 
\begin{equation}
N=\frac{\ln(\frac{\beta }{\alpha }t_{N}+1)}{\ln(1+l\beta)}. \label{N}
\end{equation}
Note that eq.(\ref{N}) holds even for
the elastic limit 
\begin{equation}
N \to \frac{t_{N}}{\alpha l}  \quad \mathrm{for} \quad 
e \to 1. 
\end{equation}
Here we have used the result of eq.(\ref{hsimu}) , i.e. $h
( e) \to 0$ for $e \to 1$. 

If we assume that the inelastic Lorentz gas system
corresponds to the random walk system with the reduction of hopping
velocity, the effect of inelasticities can be absorbed in the number of
collisions $N$. 
Thus, we deduce the expression of mean square displacement of a tracer as
\begin{eqnarray}
\langle{\bf r}^{2}(t)\rangle_{e} = l^{2}N \simeq l^{2}\frac{\ln(\frac{\beta }{\alpha}t+1)}{\ln(1+l\beta)}, \label{msd}
\end{eqnarray}
where we use eq.(\ref{N}) and the suffix $e$ of the bracket emphasizes
the quantity with the restitution coefficient $e$. 
Let us evaluate some physical quantities in eq.(\ref{msd}). 
Using the values of our simulation, we obtain
$\alpha=1/\sqrt{2}U^{-1}$, $\beta\simeq 0.058\times(1-e)d^{-1}$ from
eqs.(\ref{hsimu}) and (\ref{ab}). 
We set the mean free path of
the tracer as $l \simeq 13.5d$ instead of the theoretical mean free path $l=12.5d$. 
 
Substituting these physical quantities and eq.(\ref{hsimu}) into
eq.(\ref{msd}), we obtain the mean square
displacement as a function of time and restitution
coefficient. 
Figure \ref{msdtheory} shows $\langle{\bf r}^{2}(t)\rangle_{e} \times\ln(1+l\beta)d^{-2}$ versus $(1-e)tUd^{-1}$ plots for
$e=0.4, 0.6, 0.7$ and $0.9$. 
All the results of our
simulation shown as the symbols are on the same curve. 
Our results clearly demonstrate that the tracer is
localized in the case $e < 1$ and that the mean
square displacement seems to obey a power of the
logarithmic function of time. 
It should be emphasized that 
eq.(\ref{msd}), shown in Fig.\ref{msdtheory}, becomes identical with the results of our simulation as time goes on. 
The reason why eq.(\ref{msd}) deviates a little from the data of our
simulation in the early time region in Fig.\ref{msdtheory} is that few
collisions have happened in this region. 

\subsection{Position distribution}
In the inelastic Lorentz gas system, the position distribution
function of the tracer $g_{e}(r,t)$ for the restitution coefficient $e$ is scaled by the mean square
displacement $\langle{\bf r}^{2}(t)\rangle_{e}$. 
The scaled position distribution function of the tracer
$G(R)$ is defined as  
\begin{equation}
G(R) \equiv \sqrt{\langle{\bf
    r}^{2}(t)\rangle_{e}} \times g_{e}(r,t),\label{scaledposi}
\end{equation}
where $R\equiv r/\sqrt{\langle{\bf
r}^{2}(t)\rangle_{e}}$ is the scaled distance from the center of 
the system at which the tracer started. 
$G(R)$ is independent of the restitution coefficient $e$ and time. 
Let us study the scaled position distribution of the tracer
$G(R)$. 
Figure \ref{inelasticposi} shows the results of our simulation and
 analysis. 
The symbols are results of our simulation which show the scaled position
 distribution of the tracer for $e=0.4, 0.6, 0.7$ and $0.9$ at
 four different times $t=3000dU^{-1},300dU^{-1},500dU^{-1}$ and $1000dU^{-1}$, respectively. 
All the data of the scaled position distribution
 $G(R)$ are almost on the universal curve. 
It should be noted that the curve is close to a Gaussian near its peak, while
the tail seems to obey an exponential function. 
This exponential tail has been observed in a simulation of granular gas
system.\cite{Chen} 

On the other hand, the broken line and the dash-dotted line in
Fig.\ref{inelasticposi} are results of our theoretical analysis which
will be explained later. 
The broken line corresponds to the data in the tail
region of the scaled position distribution, while the dash-dotted line
corresponds to the data in the peak region of the scaled
position distribution. 
Theoretical curves well explain the result of our simulation. 
Now, let us explain how to obtain the theoretical curves. 

At first, we assume that the tracer obeys a renormalized diffusive
motion as assumed in the previous subsection. 
This assumption is validated by the derivation of the diffusion equation 
of the tracer by means of a Chapman-Enskog expansion in the density
gradient of the tracer.\cite{Chapmann,Brey3}
The time evolution of the position-velocity distribution of the tracer
$p({\bf r},{\bf v},t)$ is given by the inelastic Lorentz gas
equation (see Appendix A):  
\begin{equation}
\frac{\partial p({\bf r},{\bf v},t)}{\partial t}+{\bf v}\cdot\nabla p({\bf r},{\bf v},t)=\Lambda p({\bf r},{\bf v},t), \label{diffuse1}
\end{equation}
where the operator $\Lambda$ is defined as 
\begin{equation}
\Lambda p({\bf r},{\bf v},t)=d \rho_{0}\int_{{\bf v} \cdot {\bf \hat{\sigma}}>0} ({\bf v} \cdot {\bf \hat{\sigma}}) [\frac{p({\bf r},{\bf v}^{*},t)}{e^{2}}-p({\bf r},{\bf v},t)]  \mathrm{d}{\bf \hat{\sigma}}. \label{diffuse1.5}
\end{equation}
Integrating this equation by ${\bf v}$, we obtain
\begin{equation}
\frac{\partial n({\bf r},t)}{\partial t}+\nabla \cdot {\bf J}({\bf r},t)=0, \label{diffuse2}
\end{equation}
where the density of the tracer $n({\bf r},t)$ is defined as 
\begin{equation}
n({\bf r},t)=\int \mathrm{d}{\bf v} \mathit{p}({\bf r},{\bf v},t), \label{diffuse3}
\end{equation}
and the $\mathit{flux}$ of the tracer ${\bf J}({\bf r},t)$ is defined as  
\begin{equation}
{\bf J}({\bf r},t)=\int \mathrm{d}{\bf v} {\bf v} \mathit{p}({\bf r},{\bf v},t). \label{diffuse4}
\end{equation}
Note that the integral of the right hand side of eq.(\ref{diffuse1}) by the
velocity of the tracer becomes zero, because it denotes the effect of
change in the velocity of the tracer.      

Here we adopt the Chapman-Enskog method in which the position-velocity distribution of the tracer
$p({\bf r},{\bf v},t)$ is written as a series expansion in a formal
parameter $\epsilon$, 
\begin{equation}
p({\bf r},{\bf v},t)=p^{(0)}({\bf r},{\bf v},t)+\epsilon p^{(1)}({\bf r},{\bf v},t)+\epsilon^{2}p^{(2)}({\bf r},{\bf v},t)+\cdots. \label{diffuse5}
\end{equation}
$\epsilon$ implies an implicit $\nabla n({\bf r},t)$ which provokes
diffusion of the tracer. 
Consistently, eq.(\ref{diffuse2}) is expanded as 
\begin{equation}
\frac{\partial n({\bf r},t)}{\partial t}=\sum_{j=0}^{\infty}\epsilon^{j+1}\partial_{t}^{(j)}n({\bf r},t), \label{diffuse6}
\end{equation}
with
\begin{equation}
\partial_{t}^{(j)} n({\bf r},t)=-\nabla^{(0)} \cdot {\bf J}^{(j)}({\bf r},t), \label{diffuse7}
\end{equation}
and
\begin{equation}
{\bf J}^{(j)}({\bf r},t)=\int \mathrm{d}{\bf v} {\bf v} \mathit{p}^{(j)}({\bf r},{\bf v},t). \label{diffuse8}
\end{equation}
Note that 
\begin{equation}
\nabla =\epsilon\nabla^{(0)}+\epsilon^{2}\nabla^{(1)}+\cdots. \label{diffuse9}
\end{equation}
The zeroth order of eq.(\ref{diffuse1}), thus, becomes 
\begin{equation}
\frac{\partial T(t)}{\partial t}\frac{\partial p^{(0)}({\bf r},{\bf v},t)}{\partial T(t)}=\Lambda p^{(0)}({\bf r},{\bf v},t). \label{diffuse10}
\end{equation}
A comparison of this with eq.(\ref{inelaslorentz2}) indicates that
$p^{(0)}({\bf r},{\bf v},t)$ is proportional to $f({\bf v},t)$. 
Owing to the normalization, 
\begin{equation}
p^{(0)}({\bf r},{\bf v},t)=\frac{n({\bf r},t)}{n}f({\bf v},t). \label{diffuse11}
\end{equation}
Using this relation and eq.(\ref{diffuse8}), ${\bf J}^{(0)}({\bf
r},t)=0$, i.e. the $\mathit{flux}$ of the tracer vanishes to zeroth
order, which leads to $\partial_{t}^{(0)} n({\bf r},t)=0$. 
Additionally, from eq.(\ref{diffuse11}), $\partial p^{(0)}({\bf r},{\bf
v},t)/\partial n({\bf r},t)=f({\bf v},t)/n$. 
To the first order in the density gradient $\epsilon$,
eq.(\ref{diffuse1}) leads to
\begin{eqnarray}
(\partial_{t}^{(0)} n({\bf r},t))\frac{\partial p^{(0)}({\bf r},{\bf v},t)}{\partial n({\bf r},t)}+\frac{\partial T(t)}{\partial t}\frac{\partial p^{(1)}({\bf r},{\bf v},t)}{\partial T(t)} \nonumber \\
+({\bf v}\cdot\nabla^{(0)}n({\bf r},t))\frac{\partial p^{(0)}({\bf r},{\bf v},t)}{\partial n({\bf r},t)} 
=\Lambda p^{(1)}({\bf r},{\bf v},t). \label{diffuse12}
\end{eqnarray}
Using the results of the zeroth order and eq.(\ref{tdiff}), this equation becomes 
\begin{equation}
(\xi(t)T(t)\frac{\partial }{\partial T(t)}+\Lambda) p^{(1)}({\bf r},{\bf v},t)=\frac{({\bf v}\cdot\nabla^{(0)}n({\bf r},t))}{n}f({\bf v},t),\label{diffuse13}
\end{equation}
with $\xi(t)=2dv_{0}( t)\rho_{0}h( e)$. 
The solution $p^{(1)}({\bf r},{\bf v},t)$ is proportional to
$\nabla^{(0)}n({\bf r},t)$ :
\begin{equation}
p^{(1)}({\bf r},{\bf v},t)={\bf B}({\bf v},T(t))\cdot\nabla^{(0)}n({\bf r},t),\label{diffuse14}
\end{equation}
where the vector ${\bf B}({\bf v},T(t))$ also depends on $t$ through
$T(t)$. 
Substitution of this into eq.(\ref{diffuse13}) leads to
\begin{equation}
(\xi(t)T(t)\frac{\partial }{\partial T(t)}+\Lambda){\bf B}({\bf v},T(t))=\frac{{\bf v}}{n}f({\bf v},t).\label{diffuse15}
\end{equation}
Here, the contribution to the particle flux of first order in the
density gradient is given by 
\begin{equation}
{\bf J}^{(1)}({\bf r},t)=-D_{e}(t)\nabla^{(0)}n({\bf r},t), \label{diffuse16}
\end{equation}
with
\begin{equation}
D_{e}(t)\equiv -\frac{1}{2}\int \mathrm{d}{\bf v} {\bf v}\cdot {\bf B}({\bf v},\mathit{T}(t)). \label{diffuse17}
\end{equation}
Here we have used symmetry considerations in two-dimension. 
Using the diffusion $\mathit{coefficient}$ $D_{e}(t)$, eq.(\ref{diffuse17})
becomes  
\begin{equation}
(\xi(t)T(t)\frac{\partial }{\partial T(t)}-\nu_{D})D_{e}(t)=-\frac{T}{m},\label{diffuse18}
\end{equation}
with 
\begin{equation}
\nu_{D}\equiv-\frac{\int (\Lambda{\bf B})\cdot {\bf v}\mathrm{d}{\bf v}}{\int {\bf B}\cdot{\bf v}\mathrm{d}{\bf v}}.\label{diffuse19}
\end{equation}
Because $\xi(t) \sim T^{\frac{1}{2}}(t)$, dimensional analysis requires $D_{e}(t) \sim
T^{\frac{1}{2}}(t)$.  
Therefore, 
\begin{equation}
\frac{\partial D_{e}(t)}{\partial T(t)}=\frac{D_{e}(t)}{2T(t)}.\label{diffuse20}
\end{equation}
Substitution of this into eq.(\ref{diffuse18}) becomes 
\begin{equation}
D_{e}(t)=\frac{T(t)}{m}(\nu_{D}-\frac{\xi(t)}{2})^{-1}.\label{diffuse21}
\end{equation}
From eqs.(\ref{diffuse16}) and (\ref{diffuse7}), we obtain the diffusion
equation of the tracer, 
\begin{equation}
\partial_{t}^{(1)} n({\bf r},t)=D_{e}(t){\nabla^{(0)}}^{2}n({\bf r},t),\label{diffuse22}
\end{equation}
with the diffusion $\mathit{coefficient}$ $D_{e}(t)$ in eq.(\ref{diffuse21}). 
The expression of the diffusion $\mathit{coefficient}$ $D_{e}(t)$ is obtained
explicitly in eq.(\ref{ad7}) with the aid of eq.(\ref{sonine12}) by the
method of the Sonine expansion, though the diffusion $\mathit{coefficient}$ $D_{e}(t)$ in eq.(\ref{ad7}) gives 
a little deviated results from the results of our simulation in the
inelastic Lorentz gas system as shown in Appendix C. 
Now the random walk picture is validated by this standard method. 

Therefore, the tracer obeys a renormalized diffusive
motion; the original position distribution function of the
tracer $\tilde{g}_{e}(r,t)$ is expected to be given by 
\begin{equation}
\tilde{g}_{e}(r,t)\simeq \frac{1}{4\int^{t}_{0}D_{e}(V,t^{\prime})\mathrm{d}\mathit{t}^{\prime}} r \exp[-\frac{r^{2}}{8\int^{t}_{0}D_{e}(V,t^{\prime})\mathrm{d}\mathit{t}^{\prime}}]. \label{normg}
\end{equation}
Here we have taken into account 
the distribution of hopping $\mathit{velocity}$; we have adopted the
diffusion $\mathit{coefficient}$ as 
\begin{equation}
D_{e}(V,t^{\prime})\equiv\frac{\kappa V^{2} l}{\alpha+\beta t^{\prime}}, \label{ddif}
\end{equation}
where $\kappa$ is undetermined dimensionless coefficient which will not
be indicated explicitly, $\tau$ is time interval between two adjacent collisions, ${\bf
l}$ is the hopping vector of the tracer, and $V$ is the magnitude of the scaled hopping
$\mathit{velocity}$ defined as 
\begin{equation}
V\equiv \Bigl|\frac{{\bf l}/\tau}{v_{0}(t^{\prime})}\Bigr|.  \label{L}
\end{equation}
The expression of eq.(\ref{ddif}) is derived in Appendix D. 
Introducing $\eta$ as
\begin{equation}
\eta\equiv\frac{4\int^{t}_{0}D_{e}(V,t^{\prime})\mathrm{d}\mathit{t}^{\prime}}{V^{2}\langle{\bf r}^{2}(t)\rangle_{e}}, \label{eta}
\end{equation}
we obtain the expression of $\eta$ from eq.(\ref{msd}) as
\begin{equation}
\eta = \frac {4\kappa\ln(1+l\beta)}{l\beta}, \label{etadif}
\end{equation}
which is independent of time. 
Thus, from eqs.(\ref{normg}) and (\ref{eta}), we can derive the original 
scaled position distribution function of
the tracer $\tilde{G}(R,V)$ as 
\begin{equation}
\tilde{G}(R,V)= \frac{1}{\eta V^{2}}R\mathrm{e}^{-\frac{\mathit{R}^{\mathrm{2}}}{2\eta \mathit{V}^{\mathrm{2}}}}. \label{posivel}
\end{equation}

Here we assume the scaled hopping $\mathit{velocity}$ corresponds to the
scaled velocity of the tracer ${\bf c}\equiv {\bf v}/v_{0}(t)$. 
In our simulation, all the data of the scaled velocity distribution of the tracer $F({\bf c})$ can be fitted by the Maxwell-Boltzmann distribution
$\pi^{-1} \mathrm{e}^{-{\bf c}^{2}}$ as Fig.\ref{scalev} shows, though the high
energy tail of the tracer deviates from the Gaussian (see Appendix B).  
Therefore, we assume the distribution of the scaled hopping
$\mathit{velocity}$ is almost Gaussian. 
Now, the scaled position distribution function of the tracer 
$G(R)$ is expressed by
\begin{eqnarray}
G(R) = \int^{\infty}_{0}\frac{\mathrm{e}^{{-\mathit{V}^{\mathrm{2}}}}}{\pi}\frac{1}{\eta \mathit{V}^{\mathrm{2}}}\mathit{R} \mathrm{e}^{-\frac{\mathit{R}^{\mathrm{2}}}{2\eta \mathit{V}^{\mathrm{2}}}}2\pi \mathit{V} \mathrm{d}\mathit{V}
= \frac{2}{\eta}R\int_{0}^{\infty}\mathrm{e}^{-\mathit{p}(V,R)R^{\mathrm{2}}}\mathrm{d}\mathit{V}. \label{gdefi}
\end{eqnarray}
The second equality in eq.(\ref{gdefi}) defines $p(V,R)$ as
\begin{equation}
p(V,R)\equiv\frac{\ln V}{R^2}+\frac{V^{2}}{R^2}+\frac{1}{2\eta V^{2}}.\label{pdefi}
\end{equation}

We analyze eq.(\ref{gdefi}) by the method of steepest descent in order
to obtain the tail region of the scaled position distribution function.   
Designating $V$ which satisfies $\partial p(V,R)/\partial V=0$ as $V^{*}(R)$, 
the expression of $V^{*}(R)$ is 
\begin{equation}
V^{*}(R)=\sqrt{-\frac{1}{4}+\frac{\sqrt{\eta+8R^2}}{4\sqrt{\eta}}}.\label{csolu}
\end{equation}

By the method of steepest descent, from eq.(\ref{gdefi}) we can derive
the approximate $G(R)$, i.e.  
\begin{eqnarray}
G(R) &\simeq&  \frac{2}{\eta}R\int_{0}^{\infty}\exp\left[ -p(V^{*},R)R^{2}-\frac{(V-V^{*})^{2}}{2}\frac{\partial^{2} p(V,R)}{\partial V^{2}}\big|_{V=V^{*}} R^{2}\right]\mathrm{d}\mathit{V} \nonumber \\
&=& \frac{2}{\eta}\sqrt{\frac{\pi}{2\frac{\partial^{2} p(V,R)}{\partial V^{2} }|_{V=V^{*}}}}\mathrm{e}^{-\mathit{p}(V^{*},R)R^{\mathrm{2}}}. \label{tailg} 
\end{eqnarray}
Equation (\ref{tailg}) is reduced to
\begin{equation}
G(R)\simeq \sqrt{\frac{{\pi R}}{2\sqrt{2}\eta^{\frac{3}{2}}}}\mathrm{e}^{-\sqrt{\frac{2}{\eta}}\mathit{R}},\label{semig}
\end{equation}
in the limit of $R \to \infty$.
Thus, we have shown that the tail of $G(R)$
obeys an exponential law. 
Now we use 
\begin{equation}
G(R)= \frac{2\gamma}{\eta}\sqrt{\frac{\pi}{2\frac{\partial^{2} p(V,R)}{\partial V^{2} }|_{V=V^{*}}}}\mathrm{e}^{-\mathit{p}(V^{*},R)R^{\mathrm{2}}},\label{gammatailg} 
\end{equation}
instead of eq.(\ref{tailg}). 
Here $\gamma$ is a fitting parameter. 
We estimate $\eta\simeq 0.25$ and $\gamma\simeq 7.39$ from the comparison of
eq.(\ref{gammatailg}) and the data of our simulation on the scaled
position distribution in the tail region. 

In order to obtain the peak region of the scaled position distribution
function, we replace $V$ in eq.(\ref{posivel}) by constant $\bar{V}$. 
Thus, eq.(\ref{gdefi}) becomes
\begin{equation} 
G(R) \simeq \frac{1}{\eta \bar{V}^{2}}R\mathrm{e}^{-\frac{\mathit{R}^{\mathrm{2}}}{2\eta\bar{\mathit{V}}^{\mathrm{2}}}}. \label{peakg}
\end{equation}
Thus, we expect the Gaussian near the peak. 
Adopting the most probable $V$ as $\bar{V}$, i.e. $\bar{V}=1/\sqrt{2}$,
$\eta$ is estimated to be $0.75$. 

In Fig.\ref{inelasticposi}, eq.(\ref{gammatailg}) with $\eta= 0.25$ and
$\gamma=7.39$ is expressed by the broken
line, while eq.(\ref{peakg}) with $\eta=0.75$ and
$\bar{V}=1/\sqrt{2}$ is expressed by the dash-dotted line. 
Figure \ref{inelasticposi} shows that eq.(\ref{gammatailg}) is 
in good agreement with the simulation data of the scaled
position distribution in the tail region, while the
data of our simulation on the scaled position distribution  
in the peak region are well-fitted by 
eq.(\ref{peakg}). 

\section{Discussion}
Some approximate analytical
solutions of the inelastic Boltzmann equation obtained by the method of
the Sonine expansion, namely scaled VDF of a granular gas, are in good agreement with the results of DSMC in
which HCS holds\cite{Brey2,Brey1,Jose} or of the molecular dynamics
simulations of smooth hard inelastic disks in HCS\cite{Brito}. 
However, it has been also demonstrated that the Sonine expansion diverges if the
inelasticity is high in the inelastic Boltzmann gas system.\cite{Brito} 
Solutions by the Sonine expansion
cannot express the definite exponential form
in the high energy tail, because the
expression of eq.(\ref{sonine1}) does not have terms of odd power of
${\bf c}$, while the definite exponential form contains odd power of
${\bf c}$. 
This means that we cannot express the scaled velocity distribution of the inelastic
Boltzmann gas system of high inelasticity by eq.(\ref{sonine1}) even if
we obtain the perfect form of eq.(\ref{sonine1}). 
The reason why $a_{2}$ does not exist in the inelastic Lorentz gas
system of high inelasticity (cf. eq.(\ref{compare1})) is thought to be
the same reason mentioned above: the solution by the Sonine expansion does not have terms of odd power of
${\bf c}$ and it cannot express the scaled velocity distribution of the
inelastic Lorentz gas system of high inelasticity because its high
energy tail is definitely exponential. 
Furthermore, in both of the
inelastic Boltzmann gas system and the
inelastic Lorentz gas system, some of the scaled velocity distributions
obtained by the method of the Sonine expansion become negative for large
${\bf c}$, which is obviously an odd result. 

We assume the scaled VDF of the tracer obeys the Gaussian distribution
in $\S$ 3.3, while we have derived the non-Gaussian scaled VDF of the
tracer in Appendix B.  
Though Fig. 6 shows the scaled VDF of the tracer seems to obey the
Gaussian, its high energy tail is thought to become exponential.  
The Gaussian scaled VDF shown in Fig. 6 is rationalized by the insufficiency of the number of the samples in our simulation. 
The non-Gaussianity of the scaled VDF is not essential for the
exponential tail of the scaled position distribution function of the tracer,
while it is essential for the fact that the cooling rate is not proportional
to $1-e^2$ but close to $1-e$. 
However, the cooling rate is almost proportional to $1-e^2$ in the inelastic
Boltzmann system in which the scaled VDF is
non-Gaussian.  
The difference is considered to be due to the difference between
one-body problem and many-body problem, i.e.  
the collision relation between precollision velocity and postcollision
velocity of the inelastic Lorentz gas system is quite different from
that of the inelastic Boltzmann gas system and the mean free time of the
inelastic Lorentz gas system also shows different
dependence on time from that of the inelastic Boltzmann gas system.  

We have analyzed the spatial structure of the inelastic Lorentz gas
system by both of the method of the Sonine expansion and the random walk
model. 
The mean square displacement of the tracer by the random walk
model is identical with the results of our simulation, while the mean
square displacement by the method of the Sonine expansion does not agree
well with them. 
Furthermore, the exponential tail of the scaled
position distribution function of the tracer cannot be derived by the
method of the Sonine expansion. 
Thus, to analyze the inelastic Lorentz gas system, the random walk model
is considered to work better than the method of the Sonine expansion,
though the random walk model also has some problems, \textit{e.g.} we have used 
the mean free path of the tracer which is a little different from the theoretical mean free path. 

\section{Conclusions}
We have found that the mean square displacement of the inelastic
Lorentz gas obeys a power of the logarithmic function of time and that 
the scaled position distribution of the
inelastic Lorentz gas has an exponential tail, while the peak is the
Gaussian.  
We have analytically derived the mean square displacement as a function of time and
restitution coefficient which is identical with the data of our
simulation. 
We also analytically obtained the scaled position distribution function
the tail of which is exponential and the peak of which is the Gaussian. 

\section*{Acknowledgments}
The author deeply appreciate S. Takesue, Ooshida. T,
K. Ichiki and H. Tomita for the useful comments. 
This study is partially supported by Grant-in-Aid for Science Research
Fund from the Ministry of Education, Science and Culture (Grant
No. 11740228).  

\appendix
\section{Two-Dimensional Inelastic Lorentz Gas Equation}
We explain the derivation of two-dimensional inelastic Lorentz gas equation which can
not be derived directly from the inelastic Boltzmann equation. 

We designate the position-velocity distribution function of the tracer at time
$t$ as $p({\bf r},{\bf v},t)$. 
The change of $p({\bf r},{\bf v},t)$ during minute time $\delta t$ is expressed as 

\begin{equation}
[p({\bf r},{\bf v},t+\delta t)-p({\bf r},{\bf v},t)]\mathrm{d}{\bf r}\mathrm{d}{\bf v}= -\Gamma_{-}+\Gamma_{+}+\Gamma_{\mathit{flow}},\label{dtchange}
\end{equation}
where $\Gamma_{+}$ is the gain term of $p({\bf r},{\bf v},t)$ by collisions during the minute
time, while $\Gamma_{-}$ is the loss term of $p({\bf r},{\bf v},t)$ by collisions during the
minute time. 
$\Gamma_{\mathit{flow}}$ indicates the change in $p({\bf r},{\bf v},t)$ due to
the $\mathit{flow}$ into or out of $d{\bf r}$,
$\mathit{free}$-$\mathit{streaming}$ $\mathit{term}$, which is given by
\begin{equation}
\Gamma_{\mathit{flow}}=-\mathrm{d} {\bf r} \mathrm{d} {\bf v} \delta \mathit{t} ({\bf v}\cdot \nabla)p({\bf r},{\bf v},t).\label{gammaflow}
\end{equation}
$\Gamma_{-}$ and $\Gamma_{+}$ are defined as 
\begin{equation}
\Gamma_{-}=\int p({\bf r},{\bf v},t) \mathrm{d}{\bf r}\mathrm{d}{\bf v} \rho_{0} |{\bf v}| \delta \mathit{t} \mathrm{d}\mathit{b} ,\label{gamma-}
\end{equation}
and
\begin{equation}
\Gamma_{+}=\int p({\bf r},{\bf v}^{*},t) \mathrm{d}{\bf r} \mathrm{d}{\bf v}^{*} \rho_{0} |{\bf v}^{*}| \delta \mathit{t} \mathrm{d}\mathit{b}=\int p({\bf r},{\bf v}^{*},t) \mathrm{d}{\bf r} \frac {\mathrm{d}{\bf v}}{\mathit{e}} \rho_{0} |{\bf v}^{*}| \delta \mathit{t} \mathrm{d}\mathit{b} ,\label{gamma+}
\end{equation}
respectively. 
Here ${\bf v}^{*}$ denotes precollision velocity of the tracer,
$\rho_{0}$ is the mean number density of scatters, $e$ is the restitution coefficient, $b$ indicates the
impact parameter ( see Fig.\ref{impact} ). 
Note that we use the Jacobian $\mathrm{d}{\bf v}^{*}=\mathrm{d}{\bf v}/\mathit{e}$. 
Substitution of eqs.(\ref{gamma-}),(\ref{gamma+}) and (\ref{gammaflow}) into
eq.(\ref{dtchange}) leads to
   
\begin{equation}
\frac{\partial p({\bf r},{\bf v},t)}{\partial t}+({\bf v}\cdot \nabla)p({\bf r},{\bf v},t)= \rho_{0}\int [|{\bf v}^{*}|\frac{p({\bf r},{\bf v}^{*},t)}{e}-|{\bf v}|p({\bf r},{\bf v},t)] \mathrm{d}\mathit{b}. \label{ftdiff}
\end{equation}

It is often convenient to rewrite the integral over the impact parameter 
as the unit vector ${\bf \hat{\sigma}}$ as 
 
\begin{equation}
|{\bf v}|\mathrm{d}\mathit{b} = B(|{\bf v}|,{\bf \hat{\sigma}})\mathrm{d}{\bf \hat{\sigma}}, \label{rewrite}
\end{equation}
Here $\mathrm{d}{\bf \hat{\sigma}}=\mathrm{d}(\pi-\psi)$ and $b=d \sin(\pi-\psi)$ ( see Fig.\ref{impact}
). 
$d$ is the diameter of the tracer and the scatters. 
Thus, the expressions of $B(|{\bf v}|,{\bf \hat{\sigma}})$ and $B(|{\bf v}^{*}|,{\bf \hat{\sigma}})$ are given by 
\begin{equation}
B(|{\bf v}|,{\bf \hat{\sigma}})=|{\bf v}|d \cos(\pi-\psi) =d ({\bf v} \cdot {\bf \hat{\sigma}}), \label{bexpress}
\end{equation}
and
\begin{equation}
B(|{\bf v}^{*}|,{\bf \hat{\sigma}})=|{\bf v}^{*}|d \cos(\pi-\psi) =-d ({\bf v}^{*} \cdot {\bf \hat{\sigma}})=d \frac{({\bf v} \cdot {\bf \hat{\sigma}})}{e}, \label{bexpress2}
\end{equation}
respectively ( see Fig.\ref{impact} ). 
From eqs.(\ref{ftdiff}), (\ref{rewrite}), (\ref{bexpress}) and (\ref{bexpress2}), we obtain  
\begin{equation}
\frac{\partial p({\bf r},{\bf v},t)}{\partial t}+({\bf v}\cdot \nabla)p({\bf r},{\bf v},t)=d \rho_{0}\int_{{\bf v} \cdot {\bf \hat{\sigma}}>0} ({\bf v} \cdot {\bf \hat{\sigma}}) [\frac{p({\bf r},{\bf v}^{*},t)}{e^{2}}-p({\bf r},{\bf v},t)]  \mathrm{d}{\bf \hat{\sigma}}. \label{inelaslorentz}
\end{equation}
This is nothing but the two-dimensional inelastic Lorentz gas equation. 
Integrating this equation by ${\bf r}$, we obtain eq.(\ref{inelaslorentz2}).

\section{Cooling Rate $M_{2}$}
We derive the expression of $M_{2}$ by the method of Sonine expansion. 

From eq.(\ref{thermo}), eq.(\ref{tdiff2}) is rearranged as
\begin{equation}
\frac{\mathrm{d}\mathit{v_{\mathrm{0}}( t)}}{\mathrm{d}\mathit{t}}=-\frac{\rho_{0}d \mathit{M}_{\mathrm{2}} \mathit{v}_{\mathrm{0}}^{\mathrm{2}}(\mathit{t}) }{\mathrm{2}}.\label{rearramu2}
\end{equation}
Using eqs.(\ref{inelaslorentz2}),(\ref{tdepend}) and (\ref{rearramu2}), we obtain the scaled equation for the inelastic Lorentz gas as 
\begin{equation}
I(F)=\frac{ M_{2}}{2}( 2+{\bf c} \cdot \frac{\partial}{\partial {\bf c}}) F({\bf c}). \label{ifmu2}
\end{equation}
Integration of the product of $c^{p}$ and eq.(\ref{ifmu2}) by ${\bf c}$
makes
\begin{equation}
M_{p}=\frac{M_{2}}{2}p\langle c^{p}\rangle,\label{mueq}
\end{equation}
where $\langle\cdot\cdot\cdot\rangle$ denotes an average over $F({\bf c})$.  

In the limit of small dissipation, $F({\bf c})$ approaches a Gaussian,
i.e. $F({\bf c})\simeq \phi({\bf c})\equiv \pi^{-1} \exp(-{\bf c}^{2})$.
Therefore, in the low inelasticity region, a systematic approximation of $F({\bf c})$ may be expressed by 
the Sonine expansion\cite{Brito,Noije,Brey1,Brey2,Brill,Brey3}, i.e. 
\begin{equation}
F({\bf c})=\phi({\bf c})\{1+\sum_{p=1}^{\infty}a_{p}S_{p}({\bf c}^{2})\}.\label{sonine1}
\end{equation}
The first few two-dimensional Sonine polynomials are 
\begin{eqnarray}
S_{0}(x)&=&1 \\ 
S_{1}(x)&=&-x+1 \label{sonine1.5} \\
S_{2}(x)&=&\frac{1}{2}x^{2}-2x+1.\label{sonine2}
\end{eqnarray}
It should be noted that Sonine polynomials satisfy the orthogonality relations 
\begin{equation}
\int \mathrm{d}{\bf c} \phi({\bf c}) \mathit{S_{p}}({\bf c}^{\mathrm{2}})\mathit{S}_{p^{\prime}}({\bf c}^{\mathrm{2}})=\delta_{pp^{\prime}},\label{sonine3}
\end{equation}
where $\delta_{pp^{\prime}}$ is the Kronecker delta.
From these orthogonality relations, we have
\begin{equation}
a_{p}=\int \mathrm{d}{\bf c} \mathit{S_{\mathit{p}}}({\bf c}^{\mathrm{2}})\mathit{F}({\bf c})=\langle S_{p}({\bf c}^{\mathrm{2}})\rangle.\label{sonine4}
\end{equation}
From eqs.(\ref{sonine1.5}), (\ref{sonine2}) and (\ref{sonine4}), $a_{1}$
and $a_{2}$ is given as
\begin{eqnarray}
a_{1}&=&-\langle{\bf c}^{2}\rangle+1=0 \label{sonine4.5}\\
a_{2}&=&\frac{\langle{\bf c}^{4}\rangle}{2}-2\langle{\bf c}^{2}\rangle+1=\frac{\langle{\bf c}^{4}\rangle}{2}-1,\label{sonine5}
\end{eqnarray}
respectively. 
Here we have used the fact that substitution of $p=2$ into eq.(\ref{mueq}) leads to $\langle{\bf
c}^{2}\rangle=1$.  
Substitution of $p=4$ into eq.(\ref{mueq}) leading to 
\begin{equation}
M_{4}=2 M_{2}\langle{\bf c}^{4}\rangle,\label{sonine6}
\end{equation}
we obtain
\begin{equation}
4 M_{2}(a_{2}+1)-M_{4}=0.\label{sonine7}
\end{equation}

On the other hand, the moments of the dimensionless
collision integral $M_{p}$ is calculated as 
\begin{equation}
M_{p} = \int \mathrm{d}{\bf c} \int_{{\bf c} \cdot {\bf \hat{\sigma}}>0} \mathrm{d}{\bf \hat{\sigma}} ({\bf c} \cdot {\bf \hat{\sigma}}) \mathit{F}({\bf c}) (c^{p}-c^{\prime p}),
\label{sonine8} 
\end{equation}
where $c^{\prime}$ indicates the postcollision velocity of the tracer. 
From eq.(\ref{sonine7}), we need the expression of
$M_{2}$ and $M_{4}$ to obtain $a_{2}$. 
In general, it is impossible to obtain such the moments without the
complete information of $F({\bf c})$. 
Thus, we adopt an approximate method to obtain a closed relation for
$M_{2}$, $M_{4}$ and $a_{2}$, i.e. we approximate the scaled velocity
distribution function as $F({\bf c})=\phi({\bf c})\{1+a_{2}S_{2}({\bf
c}^{2})\}$ using the expression of eq.(\ref{sonine1}) with $a_{p}=0$ for
$p\geq 3$.  
Adopting such the approximation, the second moment is calculated as 
\begin{equation}
M_{2} = \sqrt{\pi}(1-e^{2})(1+\frac{3}{8}a_{2}). \label{sonine9}
\end{equation}
The forth moment is calculated similarly as
\begin{equation}
M_{4}=\sqrt{\pi}(1-e^{2})(3+2e^{2})(1+\frac{15}{8}a_{2}). \label{sonine10}
\end{equation}
Inserting these moments into eq.(\ref{sonine7}), we obtain
\begin{equation}
a_{2}=\frac{1}{24}(1+30e^{2}\pm\sqrt{-383+828e^{2}+900e^{4}}) \quad \mathrm{if} \quad  \mathit{e} \mathrm{\ne 1}.\label{sonine11}
\end{equation}
We adopt the stable solution of $a_{2}$\cite{Brill}, i.e.
\begin{equation}
a_{2}=\frac{1}{24}(1+30e^{2}-\sqrt{-383+828e^{2}+900e^{4}}),\label{sonine12}
\end{equation}
for the inelastic case. 
Another solution of $a_{2}$ in eq.(\ref{sonine11}) is unstable. 

Here we discuss the stability of the coefficient $a_{2}$ in
eq.(\ref{sonine11}). 
To analyze the stability of $a_{2}$ we write eq.(\ref{tdepend}) in a
more general form 
\begin{equation}
f({\bf v},t)=\frac{n}{v_{0}^{2}(t)} F({\bf c}, t),  \label{ac1}
\end{equation}
which leads to 
\begin{equation}
\frac{1}{d\rho_{0}v_{0}(t)}\frac{\partial \langle{\bf c}^{p}\rangle}{\partial t}-\frac{M_{2}}{2}p \langle{\bf c}^{p}\rangle+M_{p}=0,  \label{ac2}
\end{equation}
in essentially the same way as eq.(\ref{ifmu2}). 
Using $F({\bf c}, t)=\phi({\bf c})\{1+a_{2}(t)S_{2}({\bf c}^{2})\}$ and
$\langle{\bf c}^{2}\rangle =1$ which is bound,  
substitution of $p=4$ into eq.(\ref{ac2}) leads to 
\begin{equation}
\frac{\partial a_{2}(t)}{\partial t}=\frac{d \rho_{0}v_{0}(t)}{2}\{4M_{2}(a_{2}(t)+1)-M_{4}\},    \label{ac3}
\end{equation} 
for the coefficient $a_{2}(t)$. 
To discuss the linear stability of the coefficient $a_{2}(t)$, we divide 
it into the $equilibrium$ part $\bar{a}_{2}$ and the time-dependent part
$\delta a_{2}(t)$, i.e.   
\begin{equation}
a_{2}(t)=\bar{a}_{2}+\delta a_{2}(t).    \label{ac4}
\end{equation} 
Note that $\bar{a}_{2}$ is $a_{2}$ in eq.(\ref{sonine12}).  
Substituting eq.(\ref{ac4}) into eq.(\ref{ac3}) and ignoring the higher
order $O(\delta a_{2}^{2}(t))$ terms, we obtain 
\begin{equation}
\frac{\partial \delta a_{2}(t)}{\partial t}=\frac{d \rho_{0}}{2}\sqrt{\pi}(1-e^{2})H(e)v_{0}(t)\delta a_{2}(t),    \label{ac5}
\end{equation}
which can be solved for $a_{2}(t)$ as 
\begin{equation}
\delta a_{2}(t)=\delta a_{2}(0)\exp[\frac{d \rho_{0}}{2}\sqrt{\pi}(1-e^{2})H(e)\int^{t}_{0} v_{0}(t)\mathrm{d}\mathit{t}],     \label{ac6}
\end{equation}
where 
\begin{equation}
H(e)\equiv 3\bar{a}_{2}-(\frac{15}{4}e^{2}+\frac{1}{8}). \label{ac7}
\end{equation}
Thus, the stability of the coefficient $a_{2}(t)$ depends on the signs of 
$H(e)$, i.e. $a_{2}(t)$ is stable if $H(e)$ is negative. 
Figure \ref{sign} shows $H(e)$ as a function of the restitution
coefficient $e$. 
Here we substitute the expressions of $a_{2}$ in eq.(\ref{sonine11}) as
$\bar{a}_{2}$.  
From Fig. \ref{sign}, it is shown that $a_{2}$ in eq.({\ref{sonine12}}) is 
stable, while another $a_{2}$ in eq.(\ref{sonine11}) is unstable. 

It should be also mentioned that the elastic case $e=1$ is a singular
point in eq.(\ref{sonine12}), i.e. $F({\bf
c})=\phi({\bf c})\{1+a_{2}S_{2}({\bf c}^{2})\}$ does not become the
Gaussian even if $e=1$, and that eq.(\ref{sonine12}) has real values
only in the region 
\begin{equation}
\sqrt{-\frac{23}{50}+\frac{16 \sqrt{14}}{75}}\leq e \quad( < 1). \label{compare1}
\end{equation}
The reason why $a_{2}$ does not exist in the inelastic Lorentz gas
system of high inelasticity is thought to be
that the solution by the Sonine expansion does not have terms of odd power of
${\bf c}$ and it can not express the scaled velocity distribution of the
inelastic Lorentz gas system of high inelasticity by
eq.(\ref{sonine1}) even if we obtain the perfect form of
eq.(\ref{sonine1}) because its high
energy tail is definitely exponential. 
It has been also demonstrated that the Sonine expansion diverges if the
inelasticity is high in the inelastic Boltzmann gas system.\cite{Brito} 
This means that we also can not express the scaled velocity distribution of
the inelastic Boltzmann gas system of high inelasticity. 

From eqs.(\ref{sonine9}) and (\ref{sonine12}), the expression of
$M_{2}$ as a function of the restitution coefficient $e$ is given as
\begin{equation}
M_{2}=\sqrt{\pi}(1-e^{2})(\frac{65}{64}+\frac{15}{32}e^{2}-\frac{1}{64}\sqrt{-383+828e^{2}+900e^{4}}).\label{compare3}
\end{equation}
Judging from the expression of eq.(\ref{sonine9}), the reason why $M_{2}
$ is proportional not to
$1-e^2$ but to $1-e$ is due to the deviation of the scaled velocity distribution
function from the Gaussian, i.e. if
$a_{2}=0$ in $F({\bf c})=\phi({\bf c})\{1+a_{2}S_{2}({\bf c})\}$,
$M_{2}$ in eq.(\ref{sonine9}) becomes proportional to $1-e^2$ and the
cooling rate of the inelastic Lorentz gas system becomes proportional to
$1-e^2$ analytically. 
Note that because we multiply $a_{2}$ by $1-e^{2}$ in eq.(\ref{sonine9}), $M_{2}$ holds even at the elastic case, i.e. $e=1$, though
eq.(\ref{sonine12}) does not hold at the elastic case. 

\section{Diffusion Coefficient $D_{e}(t)$}
We derive the expression of the diffusion $\mathit{coefficient}$ $D_{e}(t)$ in
eq.(\ref{diffuse21}) explicitly by the method of the Sonine expansion. 

Calculating eq.(\ref{inelaslorentz2}) in a similar way as eq.(\ref{sonine8}), we obtain 
\begin{equation}
\int \mathrm{d}{\bf v} \mathit{Y}({\bf v}) \Lambda X({\bf v}) = d \rho_{0}\int \mathrm{d}{\bf v} \int_{{\bf v} \cdot {\bf \hat{\sigma}}>0} \mathrm{d}{\bf \hat{\sigma}} ({\bf v} \cdot {\bf \hat{\sigma}}) \mathit{X}({\bf v}) [Y({\bf v}^{\prime})-Y({\bf v})],    \label{ad1}
\end{equation}
where ${\bf v}^{\prime}$ is the postcollision velocity of the tracer: 
\begin{equation}
{\bf v}^{\prime}={\bf v}-(1+e)({\bf v} \cdot {\bf \hat{\sigma}}){\bf \hat{\sigma}}.     \label{ad2}
\end{equation}
Here we approximate ${\bf B}({\bf v},T(t))$ in the first Sonine approximation 
as
\begin{equation}
{\bf B}({\bf v},T(t))\sim {\bf v} f_{a_{2}}({\bf v},t),     \label{ad3}
\end{equation}
where the velocity distribution function of the tracer $f({\bf v},t)$ is 
approximated as 
\begin{eqnarray}
f({\bf v},t) &\simeq& f_{a_{2}}({\bf v},t) \label{ad4} \\
 &\equiv &\frac{m}{2\pi T(t)} \exp[{-\frac{m}{2T(t)}{\bf v}^{2}}]\{1+a_{2}(1-\frac{m}{T(t)}{\bf v}^{2}+\frac{m^{2}}{8T^{2}(t)}{\bf v}^{4})\} . \nonumber\\
 \label{ad5}
\end{eqnarray}
Use of eq.(\ref{ad1}) into eq.(\ref{diffuse19}) leads to 
\begin{equation}
\nu_{D}= \frac{d\rho_{0}(1+e)}{2}\sqrt{\frac{2\pi T(t)}{m}}(2+\frac{3}{4}a_{2}).\label{ad6}
\end{equation}
Substitution of this expression of $\nu_{D}$ into eq.(\ref{diffuse21})
leads to 
\begin{equation}
D_{e}(t)=\sqrt{\frac{2T(t)}{\pi m}}\frac{1}{d\rho_{0}}\frac{1}{(e+1)^{2}(1+\frac{3}{8}a_{2})}. \label{ad7}
\end{equation}
This is the the expression of the diffusion $\mathit{coefficient}$ $D_{e}(t)$ in
eq.(\ref{diffuse21}) by the method of the Sonine expansion. 

However, the diffusion $\mathit{coefficient}$ $D_{e}(t)$ by the method of the
Sonine expansion does not work in the Lorentz gas system as follows.  
From the definition of the diffusion $\mathit{coefficient}$:
\begin{equation}
\langle{\bf r}^{2}(t)\rangle_{e}=4\int^{t}_{0}D_{e}(t) \mathrm{d}\mathit{t}^{\prime},  \label{ad8}
\end{equation}
substitution of eq.(\ref{ad7}) into this leads to 
the mean square displacement of the tracer by the method of the Sonine
expansion as
\begin{equation}
\langle{\bf r}^{2}(t)\rangle_{e}=\frac{8}{\pi d^{2}\rho_{0}^{2}}\frac{1}{(1+e)^{2}(1-e^{2})(1+\frac{3}{8}a_{2})^{2}}\ln[1+\sqrt{\frac{\pi}{2}}d \rho_{0}(1-e^{2})(1+\frac{3}{8}a_{2})t].  \label{ad8}
\end{equation}
Thus, we have derived the logarithmic time dependence of the mean
square displacement in terms of a standard method of gas kinetics.  

Figure \ref{soninediffuse} shows $\langle{\bf
r}^{2}(t)\rangle_{e}\times (1+e)^{2}(1-e^{2})(1+\frac{3}{8}a_{2})^{2}$
versus $(1-e^{2})(1+\frac{3}{8}a_{2})t$ plots with $a_{2}$ in
eq.(\ref{sonine12}) for $e=0.6, 0.7$ and $0.9$. 
Because $a_{2}$ for $e=0.4$ does not exist, we can not plot the results
of our simulation for $e=0.4$. 
It should be mentioned that the results of our simulation shown as the
symbols are not on a same curve. 
Equation (\ref{ad8}) shown as the broken line deviates from the results
of our simulation.  
Thus, we suggest that the diffusion $\mathit{coefficient}$ $D_{e}(t)$ by the
method of the Sonine expansion is not a good approximation in the
Lorentz gas system. 

\section{The Derivation of Diffusion Coefficient $D_{e}(V,t)$}
From the fact the velocity of the tracer ${\bf v}$ is distributed, the
diffusion $\mathit{coefficient}$ of the tracer is expected to depend on
the velocity of the tracer ${\bf v}$.  
We derive the expression of the diffusion $\mathit{coefficient}$
$D_{e}(V,t)$ in eq.(\ref{ddif}) which depends on the magnitude of the scaled hopping
$\mathit{velocity}$ $V$ as follows. 

We derive the diffusion equation which depends on the velocity of the
tracer by almost similar means which we have adopted to obtain
eq.(\ref{diffuse22}), i.e. by the means of a Chapman-Enskog expansion in
the density gradient of the tracer.\cite{Chapmann,Brey3}

Integrating eq.(\ref{diffuse1}) by ${\bf v}$ , we obtain
\begin{equation}
\int2\pi v [\frac{\partial n({\bf r},v,t)}{\partial t}+\nabla \cdot {\bf J}({\bf r},v,t)]\mathrm{d}\mathit{v}=\mathrm{0}, \label{ap1}
\end{equation}
where the density of the tracer $n({\bf r},v,t)$ is defined as 
\begin{equation}
n({\bf r},v,t)\equiv\int p({\bf r},v{\bf \hat{v}},t) \delta(|{\bf \hat{v}}|-1)\mathrm{d}{\bf \hat{v}}, \label{ap2}
\end{equation}
and the $\mathit{flux}$ of the tracer ${\bf J}({\bf r},v,t)$ is defined as
\begin{equation}
{\bf J}({\bf r},v,t)\equiv\int v{\bf \hat{v}} p({\bf r},v{\bf \hat{v}},t)\delta(|{\bf \hat{v}}|-1)\mathrm{d}{\bf \hat{v}}. \label{ap3}
\end{equation}
Note that ${\bf v}=v{\bf \hat{v}}$ with $v=|{\bf v}|$. 
Because eq.(\ref{ap1}) holds for each speed of the tracer $v$, we
obtain the equation of continuity which depends on $v$ as
\begin{equation}
\frac{\partial n({\bf r},v,t)}{\partial t}+\nabla \cdot {\bf J}({\bf r},v,t)=0.\label{ap4}
\end{equation}

Here we adopt the Chapman-Enskog method in which the position-velocity distribution of the tracer
$p({\bf r},{\bf v},t)$ is written as a series expansion in a formal
parameter $\epsilon$ as in eq.(\ref{diffuse5}). 
Equation (\ref{diffuse6}) is now replaced by 
\begin{equation}
\frac{\partial n({\bf r},v,t)}{\partial t}=\sum_{j=0}^{\infty}\epsilon^{j+1}\partial_{t}^{(j)}n({\bf r},v,t), \label{ap6}
\end{equation}
with
\begin{equation}
\partial_{t}^{(j)} n({\bf r},v,t)=-\nabla^{(0)} \cdot {\bf J}^{(j)}({\bf r},v,t),\label{ap7}
\end{equation}
and
\begin{equation}
{\bf J}^{(j)}({\bf r},v,t)=\int v{\bf \hat{v}} p^{(j)}({\bf r},v{\bf \hat{v}},t)\delta(|{\bf \hat{v}}|-1)\mathrm{d}{\bf \hat{v}}. \label{ap8}
\end{equation}

Taking into account eq.(\ref{diffuse9}), the zeroth order of
eq.(\ref{diffuse1}) becomes eq.(\ref{diffuse10}). 
Here eq.(\ref{diffuse11}) is replaced by 
\begin{equation}
p^{(0)}({\bf r},{\bf v},t)=\frac{n({\bf r},v,t)}{m(v,t)}f({\bf v},t), \label{ap11}
\end{equation}
where 
\begin{equation}
m(v,t)\equiv\int f(v{\bf \hat{v}},t) \delta(|{\bf \hat{v}}|-1)\mathrm{d}{\bf \hat{v}}. \label{ap11.5}
\end{equation}
From eqs. (\ref{ap8}) and (\ref{ap11}), 
\begin{equation}
{\bf J}^{(0)}({\bf r},v,t)=v\frac{n({\bf r},v,t)}{m(v,t)}\int {\bf \hat{v}} f(v{\bf \hat{v}},t)\delta(|{\bf \hat{v}}|-1)\mathrm{d}{\bf \hat{v}}=0, \label{ap11.6}
\end{equation}
i.e. the $\mathit{flux}$ of the tracer vanishes to zeroth
order, which leads to $\partial_{t}^{(0)} n({\bf r},v,t)=0$. 
Additionally, from eq.(\ref{ap11}), $\partial p^{(0)}({\bf r},{\bf
v},t)/\partial n({\bf r},v,t)=f({\bf v},t)/m(v,t)$. 

To the first order in the density gradient $\epsilon$,
eq.(\ref{diffuse12}) is replaced by
\begin{eqnarray}
(\partial_{t}^{(0)} n({\bf r},v,t))\frac{\partial p^{(0)}({\bf r},{\bf v},t)}{\partial n({\bf r},v,t)}+\frac{\partial T(t)}{\partial t}\frac{\partial p^{(1)}({\bf r},{\bf v},t)}{\partial T(t)} \nonumber \\
+({\bf v}\cdot\nabla^{(0)}n({\bf r},v,t))\frac{\partial p^{(0)}({\bf r},{\bf v},t)}{\partial n({\bf r},v,t)} 
=\Lambda p^{(1)}({\bf r},{\bf v},t).\label{ap12}
\end{eqnarray}
From the results of the zeroth order and eq.(\ref{tdiff}) with
$l=1/2d\rho_{0}$, i.e. 
\begin{equation}
\frac{dT( t)}{dt}=-\frac{v_{0}( t)}{l}h( e) T( t),\label{ap12.5}
\end{equation}
eq.(\ref{ap12}) becomes
\begin{equation}
(\frac{v_{0}( t)}{l}h( e)T(t)\frac{\partial }{\partial T(t)}+\Lambda) p^{(1)}({\bf r},{\bf v},t)=({\bf v}\cdot\nabla^{(0)}n({\bf r},v,t))\frac{f({\bf v},t)}{m(v,t)}.\label{ap13}
\end{equation}
Similar to eq.(\ref{diffuse14}), the solution $p^{(1)}({\bf r},{\bf v},t)$ is proportional to
$\nabla^{(0)}n({\bf r},v,t)$ :
\begin{equation}
p^{(1)}({\bf r},{\bf v},t)={\bf B}({\bf v},T(t))\cdot\nabla^{(0)}n({\bf r},v,t),\label{ap14}
\end{equation}
where the vector ${\bf B}({\bf v},T(t))$ also depends on $t$ through
$T(t)$. 
Substitution of this into eq.(\ref{ap13}) leads to
\begin{equation}
(\frac{v_{0}( t)}{l}h( e)T(t)\frac{\partial }{\partial T(t)}+\Lambda){\bf B}({\bf v},T(t))={\bf v}\frac{f({\bf v},t)}{m(v,t)}.\label{ap15}
\end{equation}

Here the contribution to the particle flux of first order in the
density gradient is given by 
\begin{equation}
{\bf J}^{(1)}({\bf r},v,t)=-D_{e}(v,t)\nabla^{(0)}n({\bf r},v,t). \label{ap16}
\end{equation}
with
\begin{equation}
D_{e}(v,t)\equiv -\frac{v}{2}\int {\bf \hat{v}}\cdot{\bf B}(v{\bf \hat{v}},T(t))\delta(|{\bf \hat{v}}|-1)\mathrm{d}{\bf \hat{v}}. \label{ap17}
\end{equation}
Note that we have used symmetry considerations in two-dimension. 
$D_{e}(v,t)$ is the diffusion $\mathit{coefficient}$ which depends on the
speed of the tracer $v$. 
 
Using this diffusion $\mathit{coefficient}$ $D_{e}(v,t)$ and eq.(\ref{ap11.5}), eq.(\ref{ap15})
becomes  
\begin{equation}
(\frac{v_{0}( t)}{l}h( e)T(t)\frac{\partial }{\partial T(t)}-\nu_{D})D_{e}(v,t)
=-\frac{v^{2}}{2}.\label{ap18}
\end{equation}
with 
\begin{equation}
\nu_{D}\equiv-\frac{\int (\Lambda{\bf B}(v{\bf \hat{v}},T(t))\cdot v{\bf \hat{v}}\delta(|{\bf \hat{v}}|-1)\mathrm{d}{\bf \hat{v}}}{\int \mathit{v}{\bf \hat{v}}\cdot{\bf B}(v{\bf \hat{v}},T(t))\delta(|{\bf \hat{v}}|-\mathrm{1})\mathrm{d}{\bf \hat{v}} }.\label{ap19}
\end{equation}
Dimensional analysis requires $D_{e}(v,t)\sim v_{0}^{-1}(t) \sim T^{-\frac{1}{2}}(t)$.  
Therefore, 
\begin{equation}
\frac{\partial D_{e}(v,t)}{\partial T(t)}=-\frac{D_{e}(v,t)}{2T(t)}.\label{ap20}
\end{equation}
Substitution of this into eq.(\ref{ap18}) becomes 
\begin{equation}
D_{e}(v,t)=\frac{v^{2}}{2}(\frac{h( e)v_{0}( t)}{2l}+\nu_{D})^{-1}\sim \frac{v^{2}l}{v_{0}(t)},\label{ap21}
\end{equation}
where we assume $\nu_{D}\sim v_{0}(t)/l$ from eq.(\ref{ap19}). 
Assuming the hopping $\mathit{velocity}$ ${\bf l}/\tau$ corresponds to
the velocity of the tracer ${\bf v}$, we obtain the diffusion
$\mathit{coefficient}$ eq.(\ref{ddif})
which depends on the magnitude of the scaled hopping
$\mathit{velocity}$ $V$ as 
\begin{equation}
D_{e}(V,t)= \kappa\frac{V^{2} l}{\alpha+\beta t},\label{ap21.5}
\end{equation}
where $\kappa$ is an unimportant dimensionless constant.

\newpage
\newpage
\newpage
\begin{figure}[htbp]
\epsfxsize=12cm
\centerline{\epsfbox{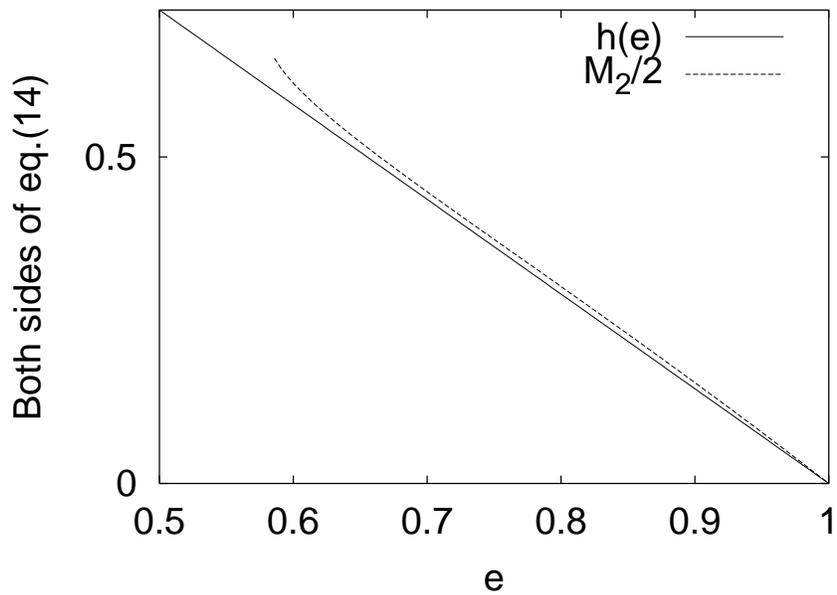}} 
\caption{$M_{2}/2$ (short-dashed line) and $h(e)$ (solid line) as a
function of the restitution coefficient $e$. $M_{2}/2$ is obtained from eq.(\ref{compare3}), while $h(e)$ is obtained from eq.(\ref{hsimu}) which has been estimated from the data of our
simulation.}
\label{hetomu2}
\end{figure}
\newpage
\begin{figure}[htbp]
\epsfxsize=12cm
\centerline{\epsfbox{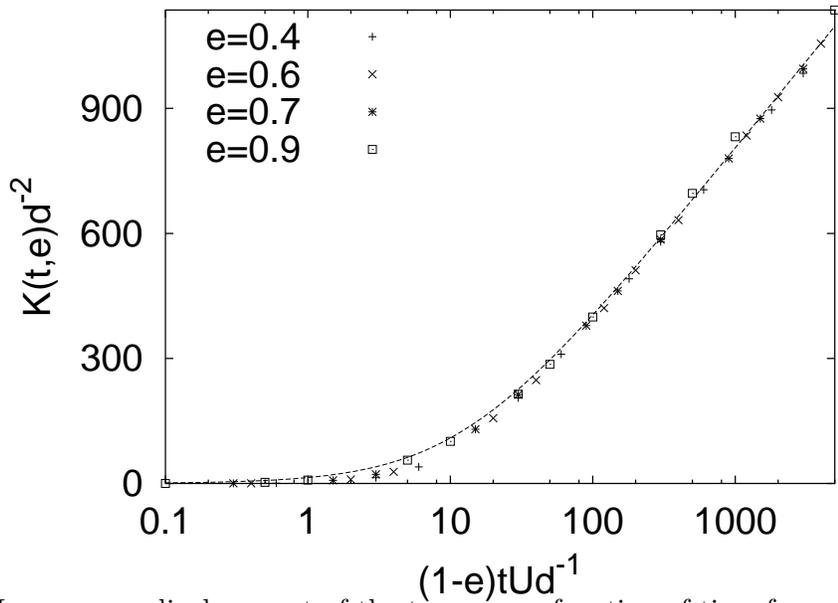}} 
\caption{Mean square displacement of the tracer as a
 function of time for several values of the restitution
 coefficient $e$. Note that $K(t,e)\equiv\langle{\bf
      r}^{2}(t)\rangle_{e} \times \ln(1+l\beta)$. The symbols are results from our
simulation. The broken line is obtained from eq.(\ref{msd}) with $l\simeq 13.5d$ and
 $\beta\simeq 0.058\times(1-e)d^{-1}$. }
\label{msdtheory}
\end{figure}
\newpage
\begin{figure}[htbp]
\epsfxsize=12cm
\centerline{\epsfbox{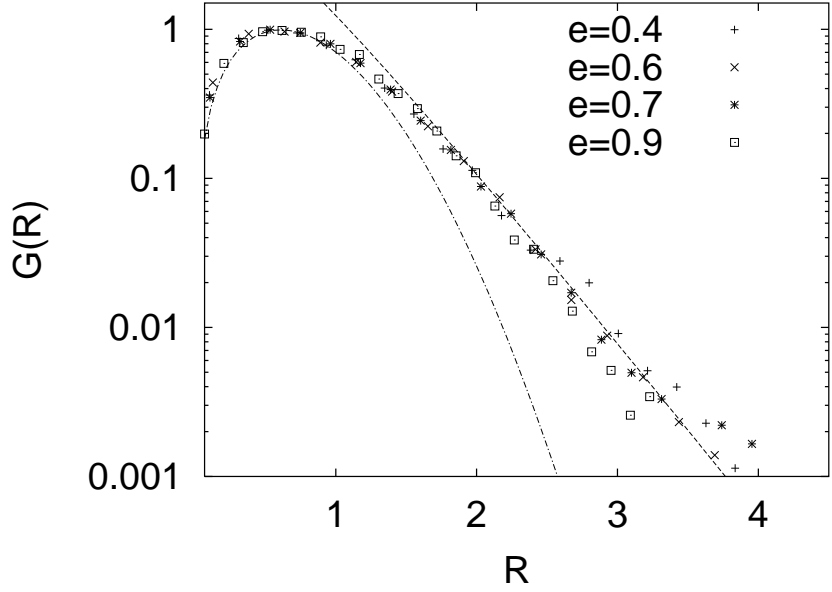}} 
\caption{Scaled position distribution 
of the tracer $G(R)=\sqrt{\langle{\bf
    r}^{2}(t)\rangle_{e}} \times g_{e}(r,t)$ as a
 function of the scaled distance 
$R\equiv r/\sqrt{\langle{\bf r}^{2}(t)\rangle_{e}}$ from the center of the
 system at which the tracer started. The symbols are the data of our simulation 
 which are plotted 
for several values of the restitution coefficient $e=0.4, 0.6, 0.7 $ and 
 $ 0.9$ at 
four different times $t=3000dU^{-1},300dU^{-1},500dU^{-1} $ and $ 1000dU^{-1}$, respectively.  The broken line and the dash-dotted line are the
 approximate solutions of $G(R)$. The
 simulation data in the tail region correspond to eq.(\ref{gammatailg}) with
 $\eta=0.25$ and $\gamma\simeq 7.39$, while the
 simulation data in the peak
 region correspond to eq.(\ref{peakg}) with $\eta=0.75$ and $\bar{V}=1/\sqrt{2}$. }
\label{inelasticposi}
\end{figure}
\newpage
\begin{figure}[htbp]
\epsfxsize=12cm
\centerline{\epsfbox{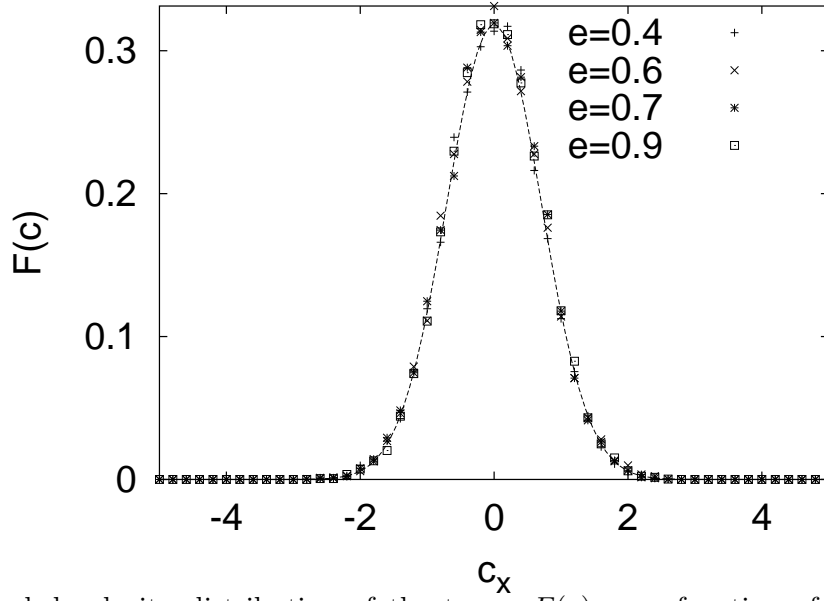}} 
\caption{Scaled velocity distribution of
the tracer $F({\bf c})$ as a function of $c_{x}$. The symbols are data of our simulation. The data are plotted 
for several values of the restitution coefficient $e=0.4, 0.6, 0.7 $ and $ 0.9$ at 
four different times $t=3000dU^{-1}, 300dU^{-1}, 500dU^{-1} $ and $ 1000dU^{-1}$, respectively. The
 broken line is the Maxwell-Boltzmann distribution $\pi^{-1}
 \mathrm{e}^{-{\bf c}^{2}}$. }
\label{scalev}
\end{figure}
\newpage
\begin{figure}[htbp]
\epsfxsize=12cm
\centerline{\epsfbox{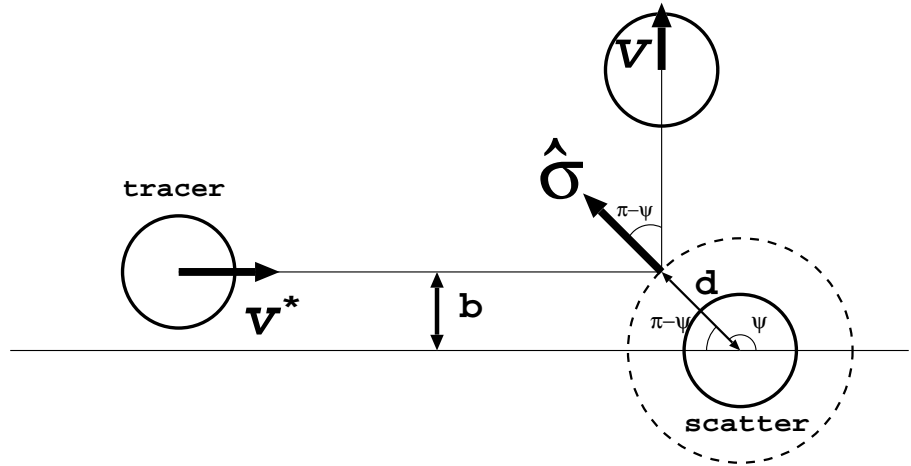}} 
\caption{Schematic description of a collision.}
\label{impact}
\end{figure}
\newpage
\begin{figure}[htbp]
\epsfxsize=12cm
\centerline{\epsfbox{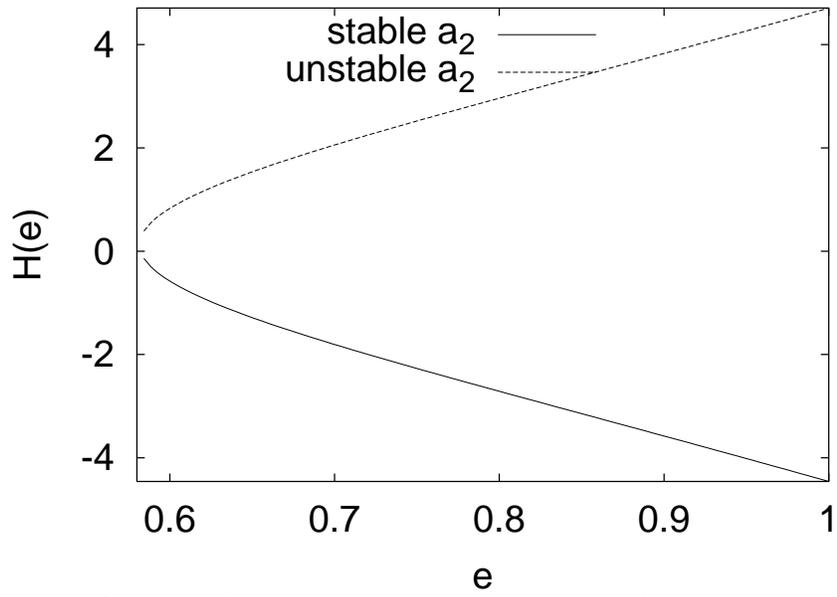}} 
\caption{$H(e)$ as a function of $e$. The solid line is $a_{2}$
 in eq.({\ref{sonine12}}), while the broken line is another $a_{2}$ in eq.(\ref{sonine11}). }
\label{sign}
\end{figure}
\newpage
\begin{figure}[htbp]
\epsfxsize=12cm
\centerline{\epsfbox{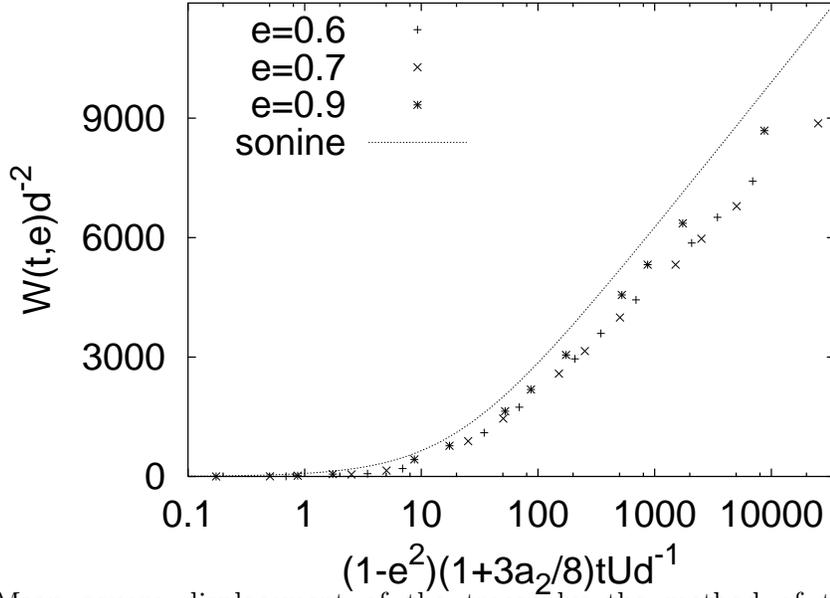}} 
\caption{Mean square displacement of the tracer by the
method of the Sonine expansion as a function of time
 for several values of the restitution coefficient $e$. Note that
 $W(t,e)\equiv \langle{\bf r}^{2}(t)\rangle_{e}\times (1+e)^{2}(1-e^{2})(1+\frac{3}{8}a_{2})^{2}$ with $a_{2}$ in
eq.(\ref{sonine12}). The symbols are results from our simulation. The
 broken line is obtained from eq.(\ref{ad8}) with $a_{2}$ in
eq.(\ref{sonine12}). }
\label{soninediffuse}
\end{figure}
\newpage

\end{document}